\documentclass[a4paper]{jpconf}
\usepackage{graphicx}
\usepackage{amsmath,amssymb}
\usepackage{cite}
\usepackage[dvips]{color}

\begin{document}
\title{Transport through a single Anderson impurity 
coupled to one normal and two superconducting leads }

\author{Akira Oguri$^1$ and Yoichi Tanaka$^2$}

\address{$^{1}$ Department of Physics, Osaka City University, 
Osaka 558-8585, Japan}

\address{$^{2}$ Condensed Matter Theory Laboratory, RIKEN, 
Saitama 351-0198, Japan}


\begin{abstract}
We study the interplay between 
the Kondo and Andreev-Josephson effects 
in a quantum dot coupled to one normal and two superconducting (SC) leads. 
In the large gap limit,
the low-energy states of this system can be described exactly by 
a local Fermi liquid for the interacting Bogoliubov particles.
The phase shift and the renormalized parameters for the Bogoliubov particles  
vary depending on the Josephson phase $\phi$ between the two SC leads. 
We explore the precise features of a crossover that 
occurs between the Kondo singlet and local Cooper-pairing states
as $\phi$ varies,  using the numerical renormalization group approach. 

\end{abstract}

\section{Introduction}

The Kondo effect in quantum dots (QD) connected to 
superconducting (SC) leads has been studied intensively 
in these years, experimentally \cite{Graber,Eich,Sand,Deacon} 
and theoretically 
\cite{Fazio,Clerk,Cuevas,Avishai,Governale,sc2,sc3,sc4,yoi_2,Hecht,Florens
}.
The competition between the strong Coulomb repulsion and 
superconductivity gives rise to the quantum phase transition (QPT)
between a nonmagnetic singlet and magnetic doublet ground states. 
It also causes the Andreev scattering  
at an interface between the QD and a normal lead,
and affects substantially transport properties near the QPT, 
or the corresponding crossover region.

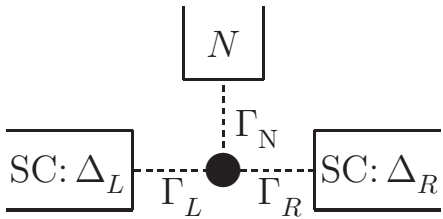
\begin{figure}[b]

\hspace{-0.8cm}
\begin{minipage}[b]{8cm}
\setlength{\unitlength}{0.75mm}

\begin{picture}(100,30)(0,0)
\thicklines

\put(17,1){\line(1,0){22}}
\put(17,15){\line(1,0){22}}
\put(39,1){\line(0,1){14}}

\put(71,1){\line(1,0){22}}
\put(71,15){\line(1,0){22}}
\put(71,1){\line(0,1){14}}

\put(48,24){\line(0,1){13}}
\put(62,24){\line(0,1){13}}
\put(48,24){\line(1,0){14}} 

\multiput(55,23.0)(0,-2){7}{\line(0,-1){1}}
\multiput(39.5,8)(2,0){7}{\line(1,0){1}}
\multiput(59.0,8)(2,0){7}{\line(-1,0){1}}

\put(55,8){\circle*{6}}

\put(57,12){\makebox(0,0)[bl]{\large $\Gamma_\mathrm{N}^{\phantom{\dagger}}$}}
\put(61,0){\makebox(0,0)[bl]{\large $\Gamma_R^{\phantom{\dagger}}$}}
\put(44,0){\makebox(0,0)[bl]{\large $\Gamma_L^{\phantom{\dagger}}$}}
\put(17.5,4){\makebox(0,0)[bl]{\large SC:$\,\Delta_L^{\phantom{\dagger}}$}}
\put(72,4){\makebox(0,0)[bl]{\large SC:$\,\Delta_R^{\phantom{\dagger}}$}}
\put(52,28){\makebox(0,0)[bl]{\large $N$}}

\end{picture}
\end{minipage}
\begin{minipage}[b]{8.5cm}
\caption{\baselineskip=0.77\baselineskip
Anderson impurity ({\large $\bullet$}) 
coupled to one normal ($N$) and two superconducting leads ($L$,$R$): 
$\Gamma_{\nu} \equiv \pi \rho \,v^{2}_{\nu}$ 
with $\rho$ the density of states 
of the lead ($\nu=L,R,N$), and 
$v_{\nu}$ the tunneling matrix element.  
The SC gaps $\Delta_{L/R} =|\Delta_{L/R}|\, e^{i\theta_{L/R}}$
give rise to the Andreev scattering and Josephson effect. 
}
\end{minipage}
%
\label{fig:system}
\end{figure}

In the present work, we consider a single QD coupled to 
one normal and two SC leads as illustrated in Fig.\ \ref{fig:system},
and study the interplay of the Kondo, Andreev, and Josephson effects,  
by using the Wilson numerical renormalization group (NRG) approach. 
We show that three phase variables, 
the phase shift $\delta$ for the interacting Bogoliubov particles, 
the Bogoliubov angle $\Theta_B$ at the interface between the QD 
and normal lead, and the Josephson phase $\phi$ between the two SC leads 
play a central role, and determine 
the low-temperature properties in the large SC gap limit.

\section{Model}

We start with a single impurity Anderson model of the form, 
\begin{align}
&H =
\xi_{d} \left(n_{d}-1\right)
+\frac{U}{2}\left(n_{d}-1\right)^2
+ 
\sum_{\nu =N,L,R} 
 \sum_{\sigma}  v_{\nu}^{}
 \left( \psi_{\nu,\sigma}^\dag d_{\sigma}^{} + 
  d_{\sigma}^{\dag} \psi_{\nu,\sigma}^{} \right) 
\, + H_\mathrm{lead}  \, +\, H_\mathrm{BCS}^{} \;, 
%
\\
& H_\mathrm{lead} =   \sum_{\nu=N,L,R}
\sum_{k,\sigma} \, \epsilon _{k}^{}\,
c_{\nu,k\sigma}^\dag c_{\nu,k\sigma}^{}, 
\qquad 
H_\mathrm{BCS}^{} \, 
=\, 
\sum_{\alpha = L,R} \sum_{k}\left(\Delta_{\alpha}\, 
c_{\alpha,k\uparrow}^\dag \,c_{\alpha,-k\downarrow}^\dag 
+ \textrm{H.c.}\right) .  
\label{H}
\end{align}
%
Here, $\xi_{d}\equiv\epsilon_{d}+U/2$, $U$  the Coulomb interaction, 
$n_{d} \equiv \sum_{\sigma}d^{\dag}_{\sigma}d^{}_{\sigma}$, and 
$d^{\dag}_{\sigma}$ ($c_{\nu,k\sigma}^\dag$) 
the creation operator for an electron with 
energy $\epsilon _{d}$ ($\epsilon _{k}$) 
and spin $\sigma$ in the quantum dot (lead).
The QD and leads are coupled via 
$v_{\nu}^{}$, and  
$\psi_{\nu,\sigma}^{} \equiv \sum_k c_{\nu,k\sigma}^{}/\sqrt{\mathcal{N}_\nu}$ 
with $\mathcal{N}_\nu$ 
the normalization factor one-particle states in the leads. 
We assume that  $\Gamma_{\nu} \equiv \pi \rho\, v_{\nu}^{2}$ 
and the density of states $\rho$  are independent 
of the frequency $\omega$.  
The complex superconducting gap  
 $\Delta_{L/R}^{} = |\Delta_{L/R}^{}| e^{i\theta_{L/R}}$ 
gives rise to the Josephson current  
between the two SC leads 
for finite $\phi \equiv \theta_{R}^{} -\theta_{L}^{}$.   
Since this Hamiltonian still has 
a number parameters to be explored,
in this report we concentrate on the half-filled case $\epsilon_d = -U/2$, 
assuming a symmetric junction with 
 $\Gamma_L^{} = \Gamma_R^{}$ $(\equiv \Gamma_S/2)$ and  
$|\Delta_L^{}| = |\Delta_R^{}|$ $(\equiv \Delta)$.  


\begin{figure}[t] 
 \leavevmode
\begin{minipage}{1\linewidth}
\includegraphics[width=0.31\linewidth]{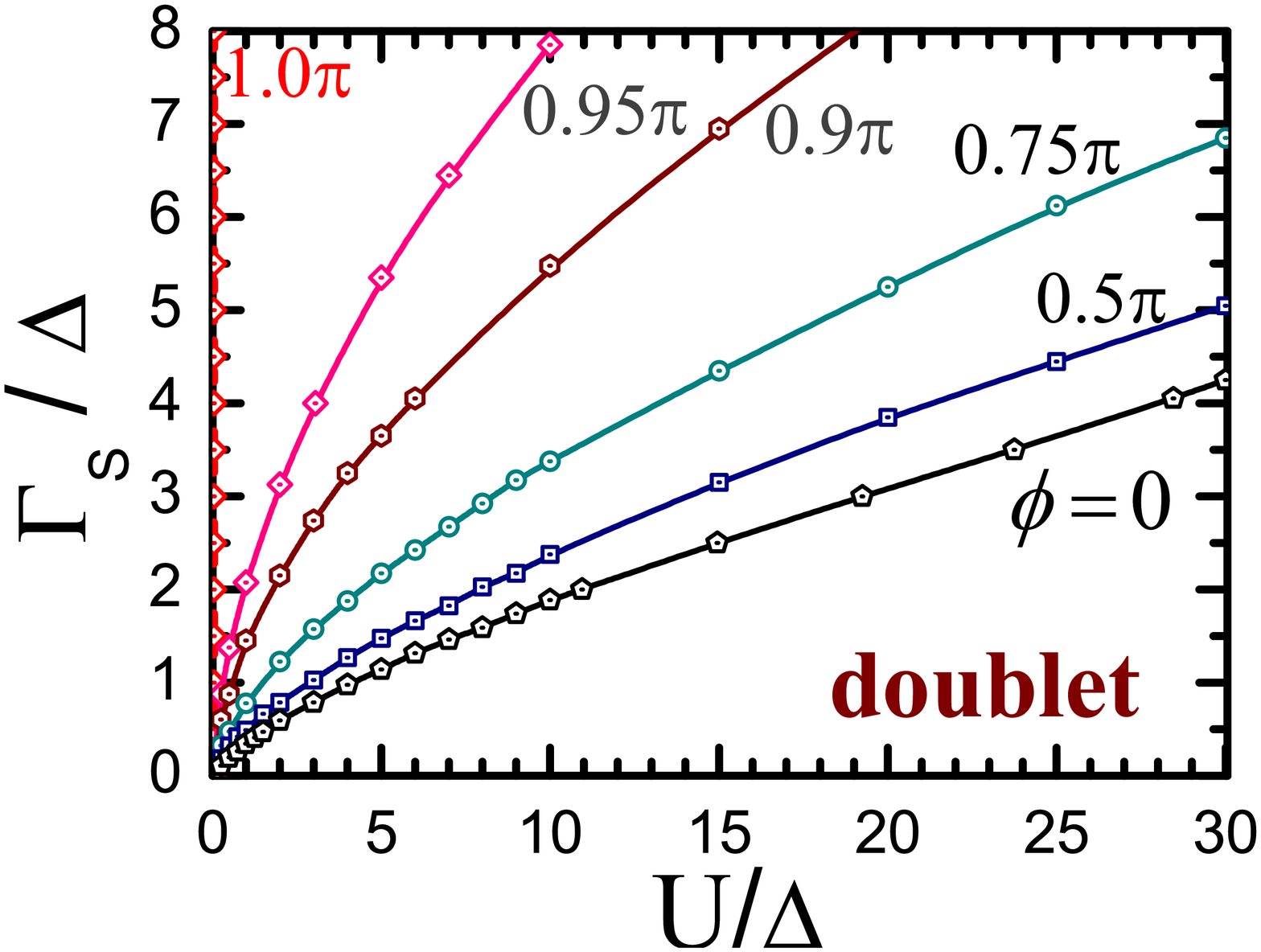}
\rule{0.01\linewidth}{0cm}
\includegraphics[width=0.31\linewidth]{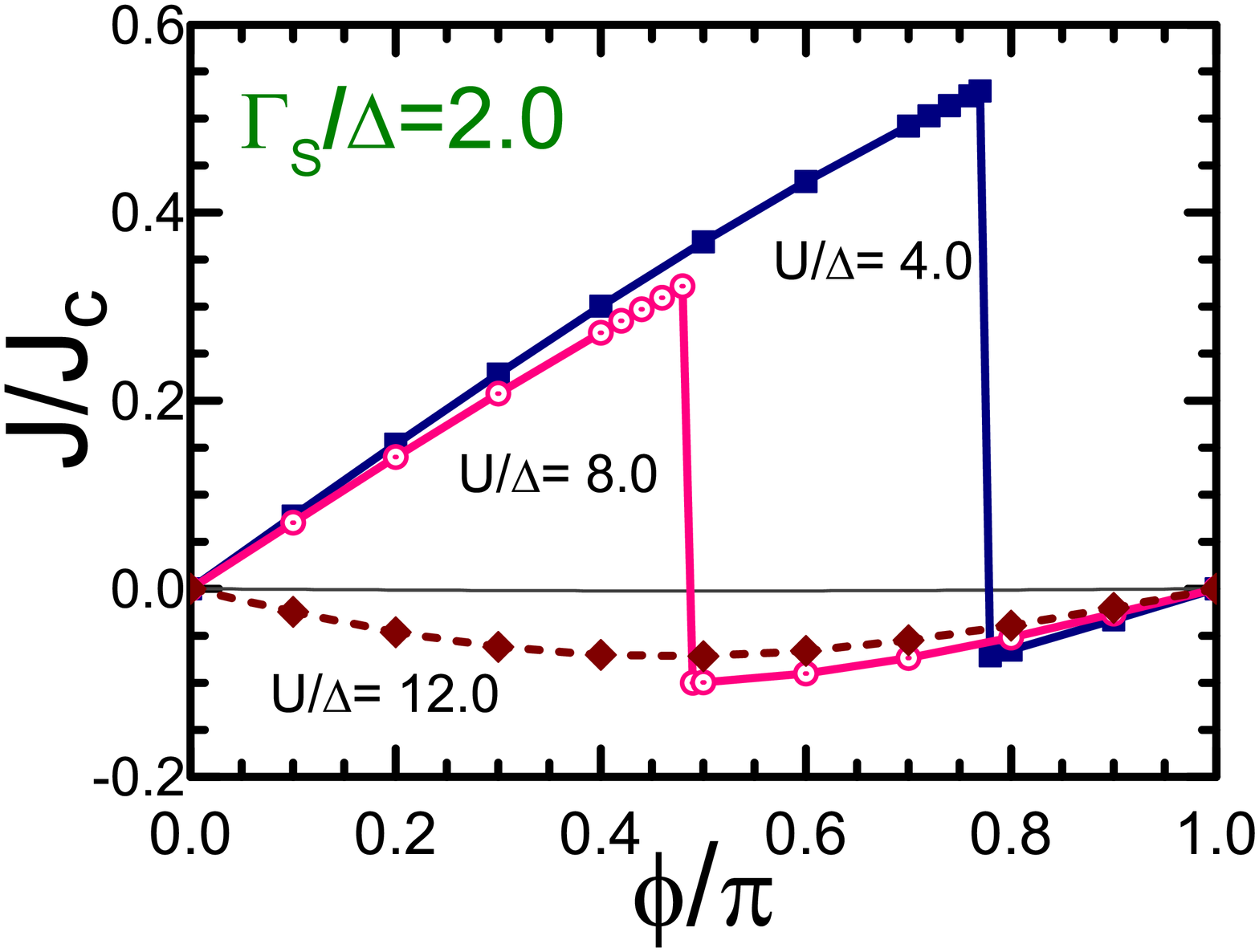}
\rule{0.01\linewidth}{0cm}
\includegraphics[width=0.31\linewidth]{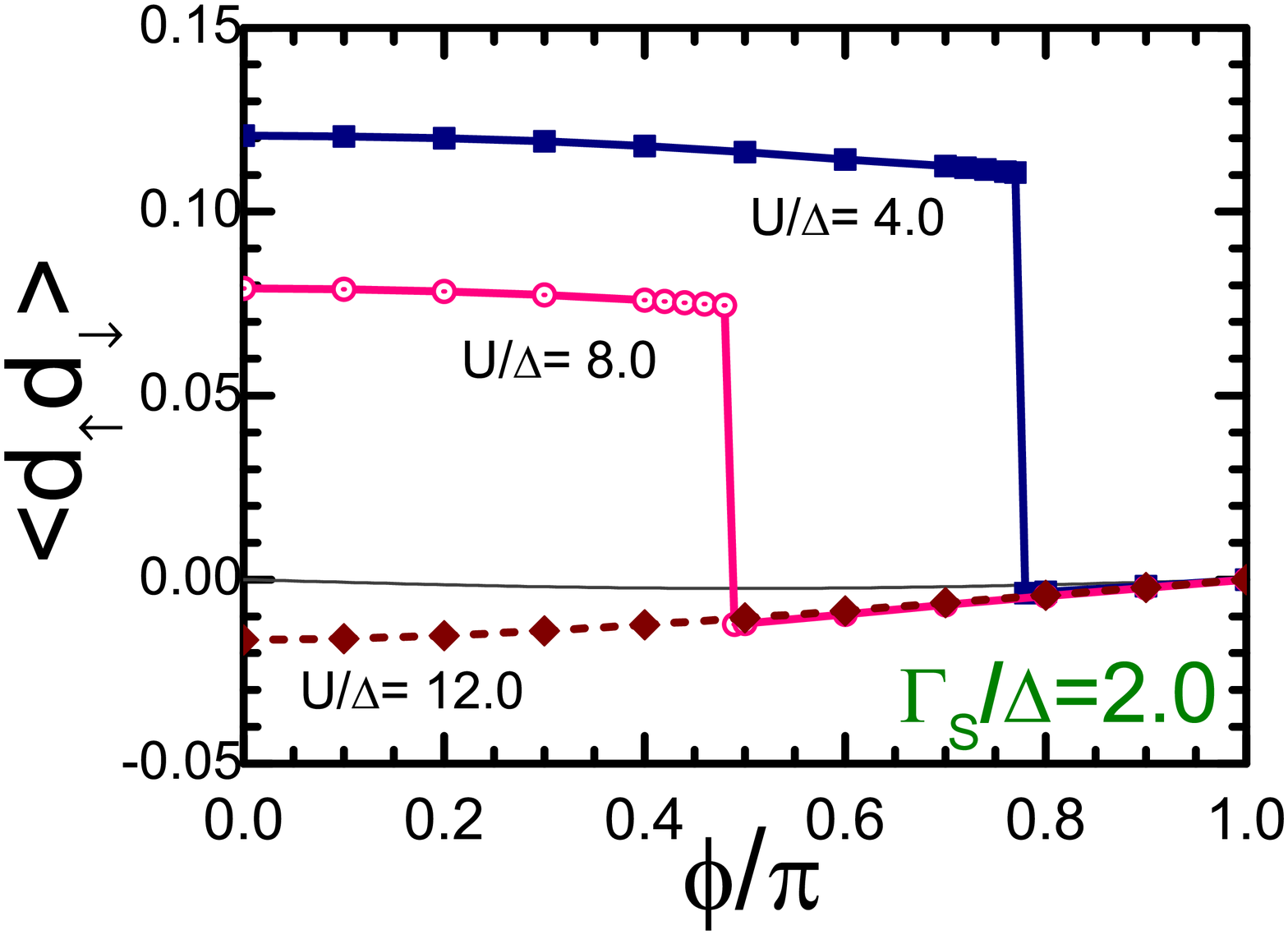}
\end{minipage}
\caption{
%
The left panel shows the phase boundary  between  
the non-magnetic singlet (upper side) and magnetic 
doulbet (lower side) ground states in the $U$ vs $\Gamma_S^{}$ plane 
for several values 
of $\phi \equiv \theta_{R}^{} -\theta_{L}^{}$. 
The middle and right panels show 
the Josephson current $J$ and $\langle d_\uparrow^{} d_\downarrow^{}  \rangle$ 
as functions of $\phi$ for $\Gamma_S^{}=2.0\Delta$  
for several $U$. 
The critical current is given by $J_C= e\Delta/\hbar\,$ for finite $\Delta$. 
} 
\label{Fig:results_2SC_terminal}
\end{figure}

\section{Two SC terminal case without the normal lead $\Gamma_N=0$}

We first of all consider a simpler $\Gamma_N=0$ case, 
where 
the QD is coupled only to the two SC leads, 
in order to see 
the competition between the Kondo and Josephson 
effects \cite{sc2,sc3,sc4,Hecht,Florens}.
Figure \ref{Fig:results_2SC_terminal} shows the NRG results for  
the ground state properties.
We see that the region of the magnetic doublet state, 
which emerges for large $U/\Delta$ and small $\Gamma_S/\Delta$, 
expands in the parameter space 
as the Josephson phase $\phi$ increases. 
This indicates that $\phi$ disturbs the Kondo screening. 
At the QPT, the Josephson current and the SC correlation 
$\langle d_\uparrow^{} d_\downarrow^{} \rangle$ 
at the impurity site vary discontinuously, 
and take small negative values 
in the magnetic doublet phase for finite $\Delta$.

\begin{figure}[t]

\begin{minipage}{1\linewidth}
\leavevmode
\includegraphics[width=0.305\linewidth]{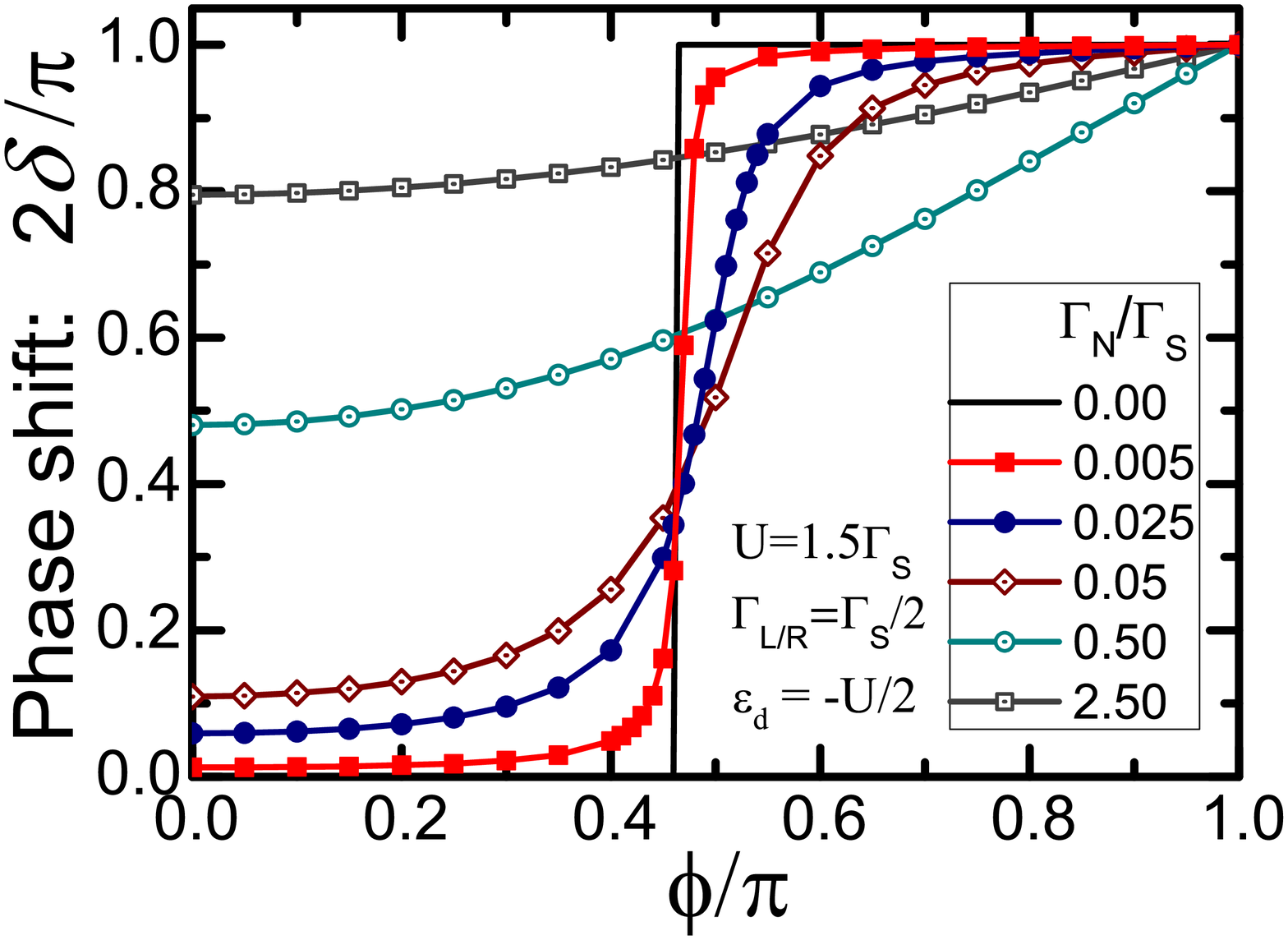}
\rule{0.01\linewidth}{0cm}
\includegraphics[width=0.31\linewidth]{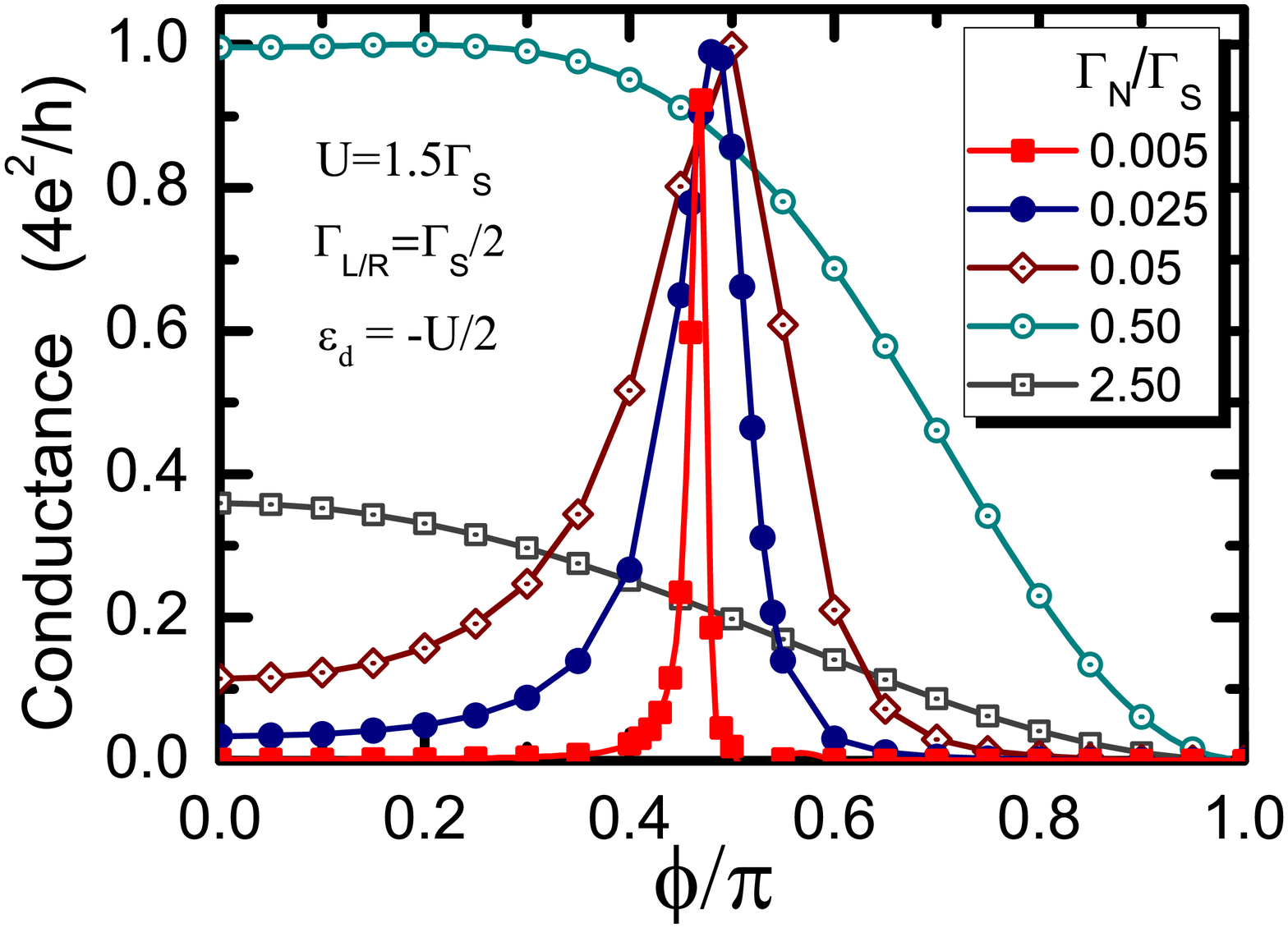}
\rule{0.01\linewidth}{0cm}
\includegraphics[width=0.31\linewidth]{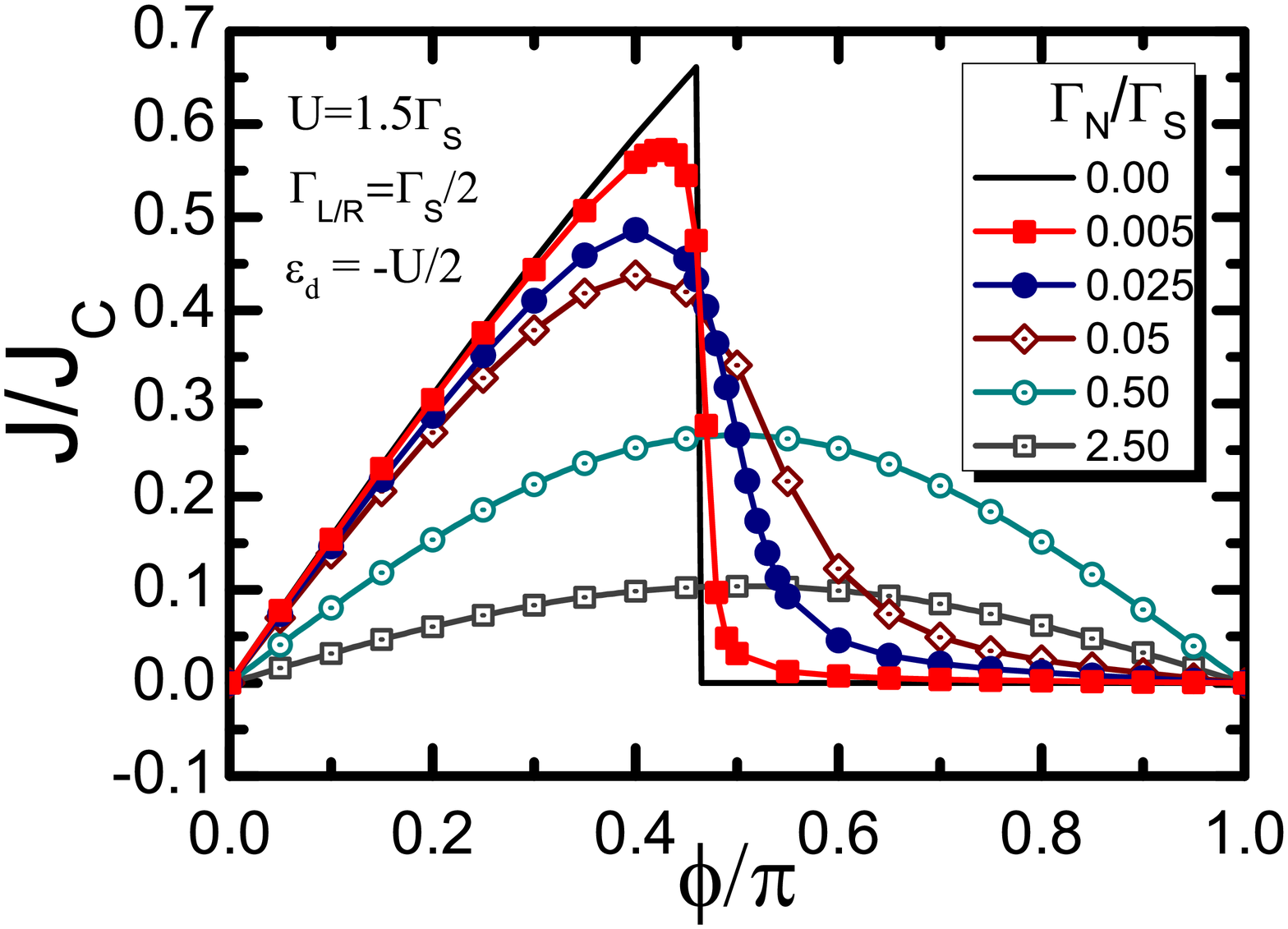}
\end{minipage}
%
\caption{
NRG results for the phase shift $\delta$ for the Bogoliubov particles (left), 
the conductance $g_{NS}^{}$ between the QD and normal lead  (middle),
and the Josephson current $J$ (right), 
are plotted vs $\phi$ for several $\Gamma_N$ in the large gap limit.  
The other parameters are chosen such that 
$U=1.5\Gamma_S$,
$\Gamma_L=\Gamma_R$ ($= \Gamma_S/2$), and $\epsilon_d=-U/2$.  
Here, $J_C = e\Gamma_S^{}/\hbar$ since $\Delta \to \infty$.}
\label{fig:YJ_cond_half}
\end{figure}

\section{Low-energy properties of the QD connected to one normal 
and  two SC leads }

We next consider the three terminal case, $\Gamma_N \neq 0$, 
with the additional normal lead.
Specifically, we concentrate 
on the $\Delta \to \infty$ limit, where $\Delta$ is much larger  
than the other energy scales $\Delta \gg 
\max(\Gamma_{S}, \Gamma_N, U, |\epsilon_d|)$,   
and the quasi-particle excitations 
in the continuum energy region above the SC gap are projected out. 
Nevertheless, the essential physics of the low-energy transport 
is still preserved in this limit, and 
a static pair potential $\Delta_d$ is induced into the impurity site 
owing to the SC proximity effects\cite{Cuevas,sc2},
\begin{align}
\Delta_d \equiv & 
\  \Gamma_R e^{i\theta_{R}^{}} + \Gamma_L e^{i\theta_{L}^{}} 
 =  |\Delta_d|\,e^{i\theta_{d}^{}}\;,
\qquad \quad  |\Delta_d| =  
\Gamma_S\,\sqrt{1 -\mathcal{T}_0 \sin^2\left(\phi/2\right)}\;,
\end{align}
where $\mathcal{T}_0 = 
{4\Gamma_R^{\phantom{0}}\Gamma_L^{\phantom{0}}}/
{(\Gamma_R^{\phantom{0}}+\Gamma_L^{\phantom{0}})^2}$.
Carrying out the Bogoliubov transform with the angle $\Theta_B$, 
\begin{align}
&\cos \Theta_B =  \frac{\xi_{d}}{E_A} , 
\qquad 
\sin \Theta_B =  \frac{|\Delta_{d}|}{E_A} ,
\qquad E_A \equiv \sqrt{\xi_{d}^2+|\Delta_{d}|^{2}} , 
\label{eq:E_d}
\end{align}
the Hamiltonian can be mapped onto the asymmetric Anderson model 
for the Bogoliubov particles, the Green's function for which 
takes the form $G(\omega)=[\omega-E_A 
+ i \Gamma_N -\Sigma(\omega)]^{-1}$ \cite{yoi_2}.
The phase shift $\delta$,  defined by $G(0)=-|G(0)|e^{i\delta}$, 
determines the ground-state properties  
\begin{align}
\langle n_{d} \rangle -1
=& \ 
\left({2\delta}/{\pi} -1 \right)
 \cos \Theta_B^{} 
\;,
\qquad \qquad 
\langle d_{\downarrow}^{} d_{\uparrow}^{} \rangle
\, = \, \frac{1}{2}
\left({2\delta}/{\pi} -1 \right)
 \,e^{i \theta_{d}^{}} \sin \Theta_B^{} \;.
\end{align}
The conductance $g_{NS}^{}$ between the QD and normal lead,
and the Josephson current $J$ between the two SC leads  
can be determined by the three phase variables 
$\delta$, $\Theta_B$ and $\phi$,  
\begin{align}
g_{NS}^{}
 = &\   
\frac{4e^2}{h}
\, \sin^2 \Theta_B^{}  \,  
\sin^2 2 \delta  , \qquad \qquad
J \, = \, 
\frac{e \Gamma_S}{\hbar}
 \,\frac{\mathcal{T}_0  \, 
\left|{2\delta}/{\pi} -1 \right|
\, \sin \Theta_B^{}\,\sin \phi 
}
{2\sqrt{1 - \mathcal{T}_0\sin^2\left(\phi/2\right)}}
\;. 
\label{eq:Condu}
\end{align}
Furthermore, the renormalization factor 
$Z= [1-\partial \Sigma/\partial\omega]^{-1}$, 
the position of the Andreev level $\widetilde{E}_A= \pm Z [E_A+\Sigma(0)]$, 
and the Wilson ratio $R$ for the Bogoliubov particles
can also be deduced from the local Fermi-liquid behavior, using the NRG.

\begin{figure}[t]

\begin{minipage}{1\linewidth}
 \leavevmode
\includegraphics[width=0.32\linewidth]{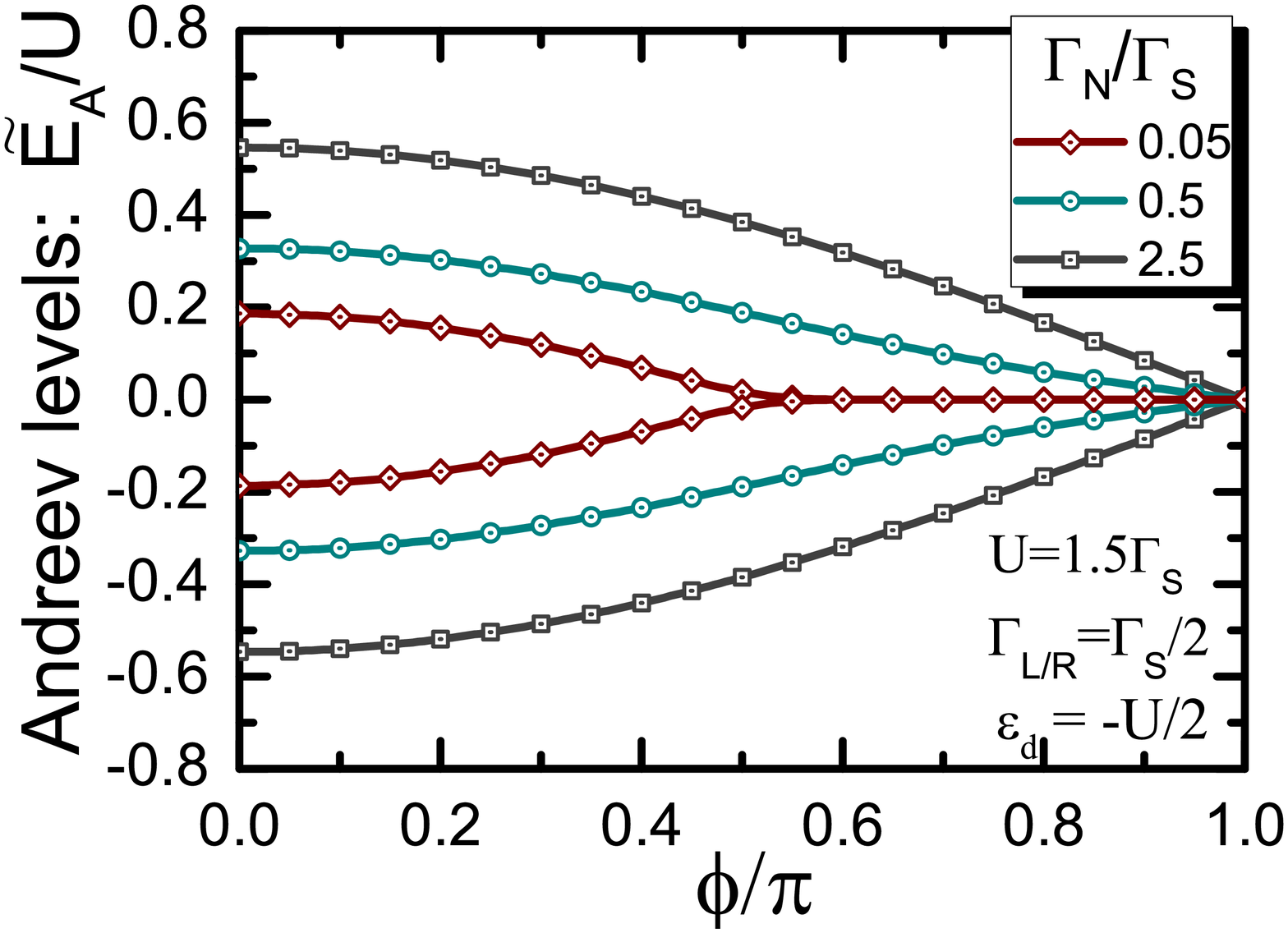}
\rule{0.018\linewidth}{0cm}
\includegraphics[width=0.31\linewidth]{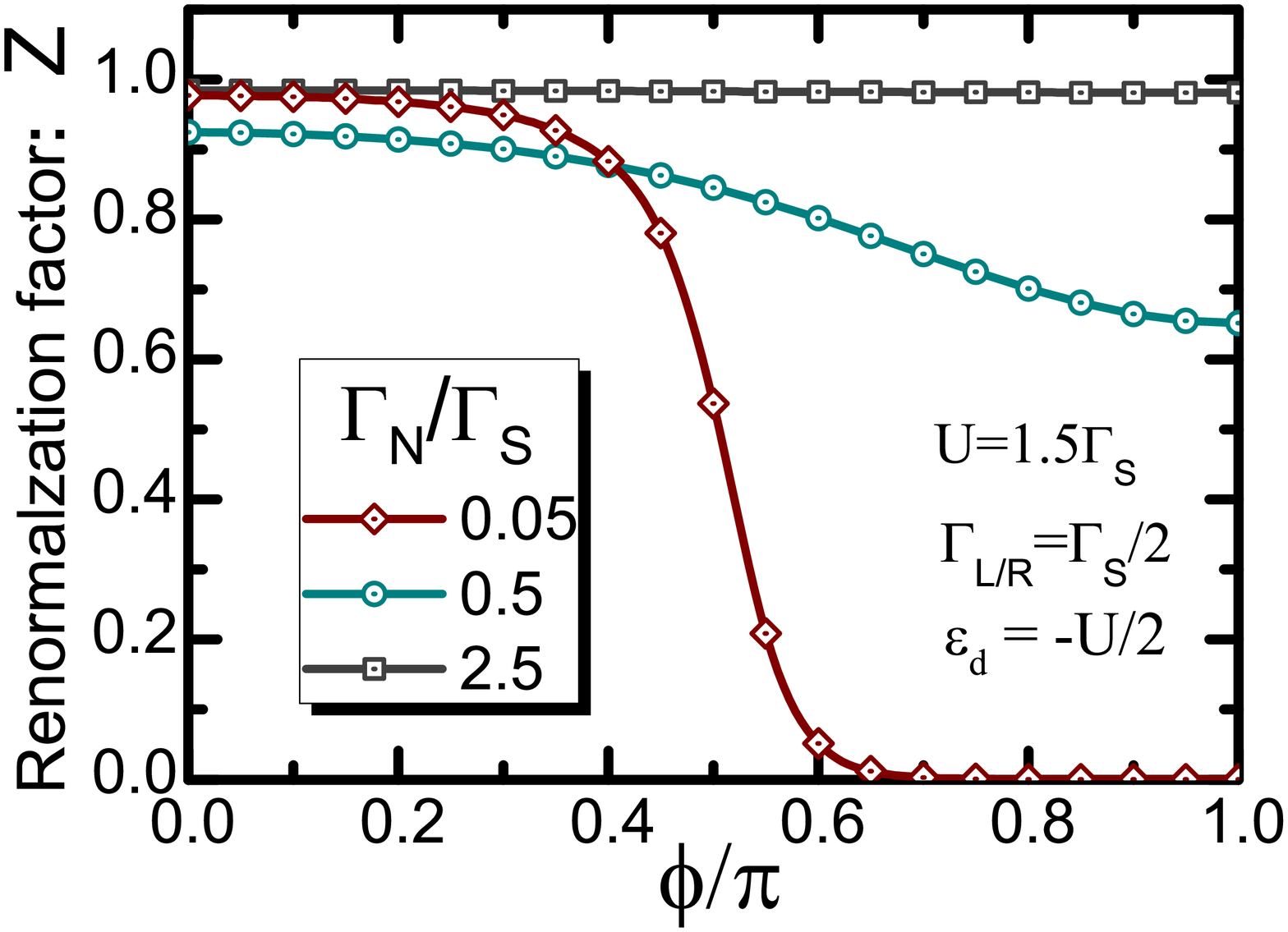}
\rule{0.007\linewidth}{0cm}
\includegraphics[width=0.32\linewidth]{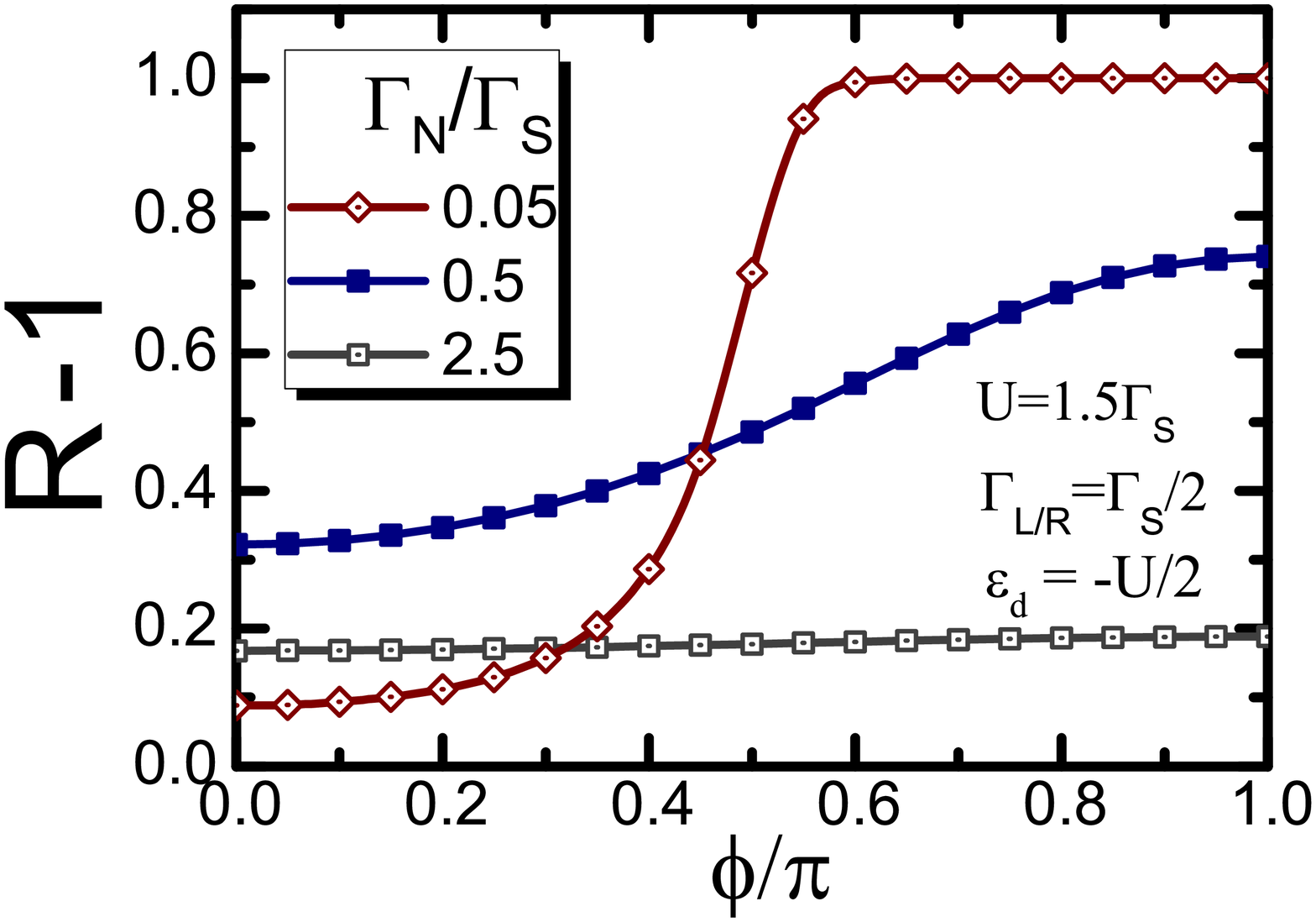}
\end{minipage}
\caption{
NRG results for the level position of renormalized Andreev resonance 
$\widetilde{E}_A$ (left), the renormalization factor $Z$ (middle)
and  $R-1$ (right), where $R$ is the Wilson ratio, 
are plotted {\it vs\/} $\phi$ in the large gap limit 
for  $U=1.5\Gamma_S$ and $\Gamma_L=\Gamma_R$ ($= \Gamma_S/2$) 
for several $\Gamma_N$. 
}
\label{fig:Renorm_half}
\end{figure}

The results are plotted  {\it vs\/} $\phi$ 
in Figs.\ \ref{fig:YJ_cond_half} 
and \ref{fig:Renorm_half} for several values of $\Gamma_N/\Gamma_S$. 
The Coulomb interaction is chosen to be $U=1.5\Gamma_S$, and 
for this value of $U$ the QPT occurs in the $\Gamma_N \to 0$ limit 
at $\phi \simeq 0.46 \pi$, where $E_A = U/2$. 
Note that $\mathcal{T}_0=1$ and  $\xi_d=0$ in the present case.
%
The gapless excitations near the normal lead 
change the sharp QPT into a continuous crossover 
between the two different singlet states. 
We see in Fig.\ \ref{fig:YJ_cond_half} that 
for $\phi \lesssim 0.46 \pi$ the ground state is a  
local Cooper pairing consisting of a linear combination of 
the empty and doubly occupied impurity states 
with a small $\delta$. On the other side, for $\phi \gtrsim 0.46 \pi$, 
the ground state is 
a Kondo singlet with $\delta \simeq \pi/2$.
At the transient region, 
the conductance $g_{NS}^{}$ has a peak, 
and the Josephson current $J$ decreases rapidly to zero.
Note that a weak current, which 
flows in the opposite direction 
as seen in Fig.\ \ref{Fig:results_2SC_terminal} 
for the magnetic doublet state with finite $\Delta$, 
is absent in the large gap limit. 
These features that are caused by the crossover  
are smeared as $\Gamma_N$ increases.

Figure \ref{fig:Renorm_half} shows the results 
for the renormalized parameters.  
We see for small $\Gamma_N$ ($=0.05\Gamma_S$) 
that the parameters are strongly renormalized 
in the Kondo-singlet region for $\phi \gtrsim 0.46 \pi$, 
where $Z \ll 1.0$,  $R \simeq 2.0$, and 
the couple of the Andreev peaks $\widetilde{E}_A$ stay 
close to the Fermi level $\omega=0$.    
In contrast,  
 in the local Cooper-pairing region for $\phi \lesssim 0.46 \pi$, 
the parameters are almost {\it not\/} renormalized 
 $Z \simeq 1.0$, $R \simeq 1.1$, and 
the Andreev peaks situate away from the Fermi level.   
When the coupling 
$\Gamma_N$ between the QD and normal lead is large,
these two singlet states cannot 
be distinguished clearly.

\section{Summary}

%

We have studied the crossover between 
the Kondo singlet  and the local Copper pairing, 
which consists of the linear combination of the empty 
and doubly occupied impurity states, 
in the quantum dot coupled to the one normal and two SC leads. 
In this three terminal geometry 
the normalized parameters, which characterize the 
Fermi-liquid behavior of the Bogoliubov particles, 
vary depending on the Josephson phase $\phi$. Therefore, 
the crossover can occur at finite Josephson phase $\phi_C^{}$. 
We calculated the phase shift $\delta$ and the renormalized 
parameters with the NRG approach,
and observed numerically that 
the Andreev conductance between the dot 
and normal lead has a peak 
near the crossover region at $\phi \simeq \phi_C$. 
The Josephson current between 
the the two SC leads decreases rapidly in the strongly 
renormalized Kondo-singlet ground state 
for $\phi \gtrsim \phi_C^{}$, where 
 the Andreev states $\widetilde{E}_A$ stay close to the Fermi level. 
In the local Cooper-paring ground state at the opposite side 
$\phi \lesssim \phi_C^{}$, 
the local Fermi-liquid parameters are not renormalized so much.

\ack
We would like to thank J.~Bauer and  N.~Kawakami for discussions.
This work is supported by the JSPS Grant-in-Aid for 
Scientific Research C (No.\ 23540375).
Y.T.\ was supported by Special Postdoctoral Researchers 
Program of RIKEN.
Numerical computation was partly carried out 
at Yukawa Institute Computer Facility.

\end{document}